\documentclass[footinbib,aps,prl,reprint,superscriptaddress,floatfix]{revtex4-2}

\usepackage{graphicx}
\usepackage{hyperref}

\begin{document}

\title{Microscopic Origin of Chiral Charge Density Wave in TiSe\texorpdfstring{$_{2}$}{2}}

\author{Hyeonjung Kim}
\affiliation{Department of Physics, Pohang University of Science and Technology (POSTECH), Pohang 37673, Republic of Korea}
\affiliation{Center for Artificial Low Dimensional Electronic Systems, Institute for Basic Science (IBS), Pohang 37673, Republic of Korea}

\author{Kyung-Hwan Jin}
\affiliation{Center for Artificial Low Dimensional Electronic Systems, Institute for Basic Science (IBS), Pohang 37673, Republic of Korea}

\author{Han Woong Yeom}
\email{yeom@postech.ac.kr}
\affiliation{Department of Physics, Pohang University of Science and Technology (POSTECH), Pohang 37673, Republic of Korea}
\affiliation{Center for Artificial Low Dimensional Electronic Systems, Institute for Basic Science (IBS), Pohang 37673, Republic of Korea}

\begin{abstract}
Chiral charge density wave (CDW) is widely observed in low dimensional systems to be entangled with various emerging phases but its microscopic origin has been elusive. We reinvestigate the representative but debated chiral CDW of TiSe$_{2}$ using scanning tunneling microscopy (STM) and density functional theory (DFT) calculations. Our STM data reveal unambiguously the chiral distortion of the topmost Se layer in domains of opposite chirality, which are interfaced with a novel domain wall. DFT calculations find the atomic structure of the chiral CDW, which has a $C2$ symmetry with the inversion and reflection symmetry broken. The chirality is determined by the helicity of Se-Ti bond distortions and their translation between neighboring layers. The present structure reproduces well the STM images with lower energy than the prevailing non-chiral $P\bar{3}c1$ structure model. Our result provides the atomistic understanding of the CDW chirality in TiSe$_{2}$, which can be referred to in a wide class of monolayer and layered materials with CDW.
\end{abstract}

\maketitle

Low-dimensional electronic systems are endowed with various many-body ground states such as superconductivity and charge density wave (CDW) and their reduced dimensionality enhances electron correlation to possibly result in exotic emerging phases. The CDW phase is a prototypical low dimensional electronic property, where the enhanced electron-phonon coupling leads to the spontaneous formation of periodic lattice distortions (PLD) and electronic renormalization such as gap opening and charge modulation at a low temperature \cite{GrunerGeorge2000Dwis}. For quite a few CDW materials, the suppression of the CDW order through doping \cite{Morosan2006,LiuY2013} and pressure \cite{Kusmartseva2009,Sipos2008,Suderow2005} has led to emerging superconductivity. However, there exist important unsolved questions on the fundamental mechanism of CDW itself and its competition with superconductivity \cite{Gruner1988}. Moreover, recent works added another complexity to CDW phenomena, namely, the existence of chiral forms of CDW in a few different materials such as transition metal dichalcogenides (TMDC) with emerging superconductivity (TiSe$_{2}$ \cite{Ishoika2010,VanWezel2010,VanWezel2011,VanWezel2012,Iavarone2012,Castellan2013,Zenker2013,Gradhand2015,Xu2020,Peng2022,Wickramaratne2022}, TaS$_{2}$ \cite{Gao2021,Guillamon_2011}, NbSe$_{2}$ \cite{Song2022}), Weyl semimetals (CoSi \cite{Li2019,Li2022}), and Kagome superconductors (KV$_{3}$Sb$_{5}$ \cite{Jiang2021,VanHeumen2021,LiHong2022}). These chiral CDW orders are suggested to be related to unconventional optical \cite{Xu2020,Wickramaratne2022}, magnetic \cite{Gradhand2015}, and/or superconducting \cite{Guillamon_2011} properties. However, the microscopic origin of chiral CDW phases has not been understood with only various phenomenological explanations suggested~ \cite{VanWezel2010,VanWezel2011,VanWezel2012}.

$1T\mathrm{-TiSe}_{2}$ can be an excellent example to show the intriguing nature of chiral CDW as the first material to be reported to exhibit chirality in its CDW \cite{Ishoika2010}.
It has a hexagonal $\mathrm{P\bar{3}m1}$ structure with Se-Ti-Se sandwiched layers in the normal state. Below $\sim$200 K, it undergoes a phase transition into a commensurate CDW phase with a periodic lattice distortion (PLD) of a $2\times2\times2$ superstructure \cite{DiSalvo1976}.
The superconductivity emerges under Cu doping \cite{Morosan2006} or high pressure \cite{Kusmartseva2009} as competing with the CDW order.
The chiral CDW order was first noticed locally by scanning tunneling microscopy (STM) and was suggested to come from the helical charge ordering between neighboring layers \cite{Ishoika2010}.
This out-of-plane charge ordering, however, needs a unit of three staked layers, which obviously does not match with the bilayer CDW order. The subsequent STM works could not find any difference in the CDW pattern between neighboring layers, which is essential in the above chiral CDW model \cite{Hildebrand2018}.
Later theoretical works introduced a hidden out-of-plane orbital order to explain phenomenologically the possible existence of the chiral CDW, which has not been verified by experiments \cite{VanWezel2010,VanWezel2011}. Very recently, an optical spectroscopy work showed that the chiral light polarization during cooling can induce a chirally-active CDW phase, suggesting strongly that the pristine CDW phase is globally a mixture of two chiral domains \cite{Xu2020}. While this work seems to support the existence of the chirality of the CDW phase itself, the microscopic origin of the chirality remains largely veiled.

In this Letter, we revisit the chirality of the CDW state of $1T\mathrm{-TiSe}_{2}$ using high-resolution STM measurements and DFT calculations.
We confirm that the Se surface layer of $1T\mathrm{-TiSe}_{2}$ at low temperature is composed of domains of distinct chirality, which appear unambiguously in STM topography and its FFT. The chiral domains are connected by a novel type of domain wall, which switches chirality and translates phase of CDW at the same time. Our DFT calculations find a lower energy chiral CDW structure, which breaks the reflection symmetry of a $P3$ structure to reduce to a $C2$ symmetry. In this model, the Se-Ti distortions for the CDW formation have a helical structure within a layer together with an interlayer translation and, thereby, fits with the bilayer out-of-plane periodicity and the previous STM works. The STM simulation based on this model agrees excellently with the experimental images. This work provides a microscopic understanding of the chiral CDW and stresses the importance of simple but large structural degree of freedom in CDW materials.

\begin{figure*}[t]
  \includegraphics{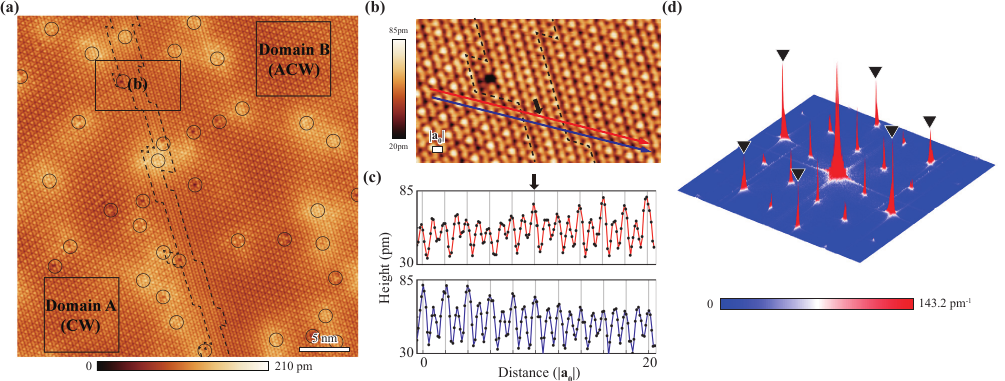}
  \caption{\textbf{(a)} Wide area topographic STM image of $1T\mathrm{-TiSe}_{2}$ ($I_{t}$ = 100 pA and $V_{s}$ = 100 mV).
  Black circles in (a) represent native defects. 
  \textbf{(b)} Enlarged image for the area within a black rectangle of (a). The CDW distortions on the topmost Se layer are shown as one bright and three less bright atoms per unit cell in (b). $2\times2$ CDW superstructures are evident in red spots in (d). Bragg peaks are represented by the black triangles. The central part of (b) shows further the longitudinal rows of atoms (between the dashed lines) without the CDW modulation, which corresponds to a translative CDW domain wall. 
  \textbf{(c)} At the domain wall, the CDW unit cell shifts along the rows as shown in the line profiles (the red and blue arrows). 
  The position of the domain wall is guided by a black arrow. $\left|\mathbf{a_{0}}\right|$ is the size of an atomic unit cell (3.54{\AA}) or a half of the CDW period.
\textbf{(d)} The fast Fourier transformation (FFT) of the STM image in (a). 
  }
  \label{fig:1}
\end{figure*}


Our experiments were performed with a commercial cryogenic STM (Unisoku, Japan) at 78 K and the mechanically cut Pt-Ir tips. The crystals were cleaved at ultra-high vacuum and applied at a bias voltage $V_{s}$ in a constant current mode with a set current $I_{t}$ to take topography images. Our DFT calculations were performed using the projected augmented plane-wave method implemented in the \textit{Vienna ab initio simulation} package \cite{Blochl1994,Kresse1996} and the generalized gradient approximation for the exchange and correlation potential \cite{Perdew1996}. Spin-orbit coupling was included. Electronic wave functions were expanded in plane waves with an energy cutoff of 520 eV. Mono- and bi-layers of $\mathrm{TiSe}_{2}$ with a $2\times2$ in-plane unit cell were considered within a supercell geometry where the vacuum spacing was 20 \AA. The residual force criterion to terminate atomic relaxation is 0.01 eV\AA$^{-1}$. We used an $11\times11\times1$ $k$-point grid to sample the entire Brillouin zone.


A high-resolution STM topographic image is shown in Fig. \hyperref[fig:1]{1(a)} for a wide area of $\mathrm{TiSe_{2}}$ at 78 K. The CDW structure in atomic scale is clear in an enlarged image shown in Fig. \hyperref[fig:1]{1(b)} as the periodically modulated contrast of protrusions, which correspond to Se atoms on the topmost layer (one bright and three less bright atoms in a $2\times2$ unit cell, see Fig. 2). The fast Fourier transformation (FFT) of the topography clearly indicates the CDW order with superstructure spots [FIG. \hyperref[fig:1]{1(d)}]. In the large area image with well-ordered CDW domains [see also FIG. S1(a) in Ref. \cite{SM2022}], one can notice quite a few points defects, which are characterized in detail in the previous STM studies \cite{Hildebrand2014,Novello2015}. 
These defects are associated with the slowly varying STM contrast, which will not be discussed here. In addition, one can also find a line defect at the center of the image. 
As depicted in the enlarged image [FIG. \hyperref[fig:1]{1(b)}], this defect shifts the phase of CDW by a single atomic unit cell in a longitudinal direction. Due to this shift, the line profiles change from the row of bright (less bright) protrusions to less bright (bright) ones when it crosses the defect [FIG. \hyperref[fig:1]{1(c)}]. That is, this is a translative CDW domain wall (see also FIG. S1(b) in Ref. \cite{SM2022}. The domain walls are also measured in other adjacent chiral domains.). The domain wall is, however, rather counterintuitive since there is no CDW translation (phase shift) in the transverse direction, that is, perpendicular to the domain wall. A few previous STM studies also noticed similar domain walls but the nature of the domain wall has not been sufficiently clear \cite{Iavarone2012,Novello2017,Yan2017,Hildebrand2016,Hildebrand2017,Lee2021}, which is closely related to the chirality as explained below.

\begin{figure*}[t]
  \includegraphics{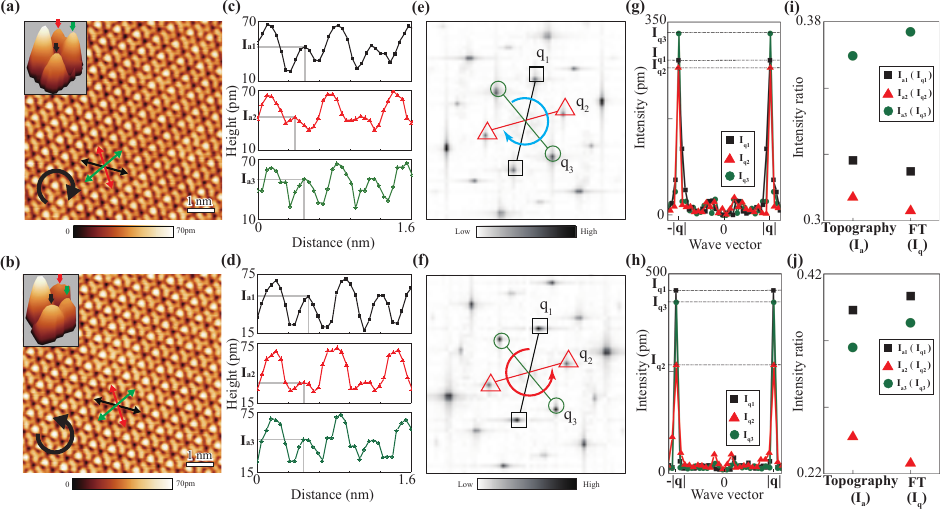}
  \caption{\textbf{(a),(b)} The enlarged images in domain A and B of Fig. \hyperref[fig:1]{1(a)} with the close-ups of their CDW unit cells in the insets (see also the unit cells of topography in FIG. S2(b) of Ref. \cite{SM2022}). \textbf{(c),(d)} Height profiles along crystallographic directions ($a_{1}$, $a_{2}$, and $a_{3}$) shown in (a) and (b) with colored arrows, respectively. \textbf{(e),(f)} FFT images of topographies in (a) and (b), respectively. \textbf{(g),(f)} Intensity profiles of the FFTs in (e) and (f), respectively along three directions indicated by colored lines ($q_{1}$, $q_{2}$, and $q_{3}$ for $a_{1}$, $a_{2}$, and $a_{3}$ in real space, respectively). \textbf{(i),(j)} The height of minor protrusions (corresponding to the less bright atoms) for three directions (denoted as $I_{a1}$, $I_{a2}$, and $I_{a3}$); The FFT intensities of the CDW superstructure spots (symbols in solid lines in (e) and (f)) (named $I_{q1}$, $I_{q2}$, and $I_{q3}$)
}
  \label{fig:2}
\end{figure*}

The role of the domain wall can be made clearer by comparing the detailed atomic scale STM images of neighboring domains. Due to the very low density of the defects and the relatively large defect-free area in our measurements, the pristine CDW unit cells of each domain can be precisely compared as presented in Fig. \hyperref[fig:2]{2}.
We select two representative areas of domain A and B on different sides of line defects, which are sufficiently far from any defect as shown in Figs. \hyperref[fig:2]{2(a)} and \hyperref[fig:2]{2(b)}. The uniformity of CDW unit cells is apparent in these uncorrected images, which are fully reflected in the sharp spots in their FFT [Figs. \hyperref[fig:2]{2(e)} and \hyperref[fig:2]{2(f)}]. We plot the line profiles along three equivalent crystallographic directions ($a_{1}$, $a_{2}$, and $a_{3}$ along the rows of Se atomic protrusions) in Figs. \hyperref[fig:2]{2(c)} and \hyperref[fig:2]{2(d)}. These profiles reveal that the protrusions are not equivalent along the directions; the three less bright protrusions within a single CDW unit cell have different contrasts. This indicates that the heights of three Se atoms in a unit cell are different to break the $P3$ rotational symmetry.

Moreover, the hierarchy of these heights is reversed in domain A and B (green-black-red in A but black-green-red atoms in B for the order of their heights). That is, the CDW structures of domain A and B are different with two distinct chirality indicating a lower symmetry than expected.
One can assure that there exists chiral structural modulation at least in the topmost Se layer. This chiral modulation is uniform over the entire CDW unit cells of each domain as measured in the images of Fig. \hyperref[fig:2]{2} and for many other defect-free areas of different domains.
This uniformity can be clearly verified by the FFT images. The three superstructure peaks for the CDW modulation along three crystallographic directions ($q_{1}$, $q_{2}$, and $q_{3}$) have different intensities and their intensity hierarchy is reserved between domain A and B.
We will call these domains clockwise (CW) and anticlockwise (ACW) domains according to the helicity of height modulations defined in the figure.
This result unambiguously verify the chiral CDW in agreement with the original claim \cite{Ishoika2010,Iavarone2012}. As shown above, the chiral distortion is not related to any point defects. Furthermore, the chiral direction is reversed systematically for all domain walls observed without a single exception. That is, it cannot be due to any artifact of STM tips and imaging conditions.

\begin{figure}[tb!]
  \includegraphics{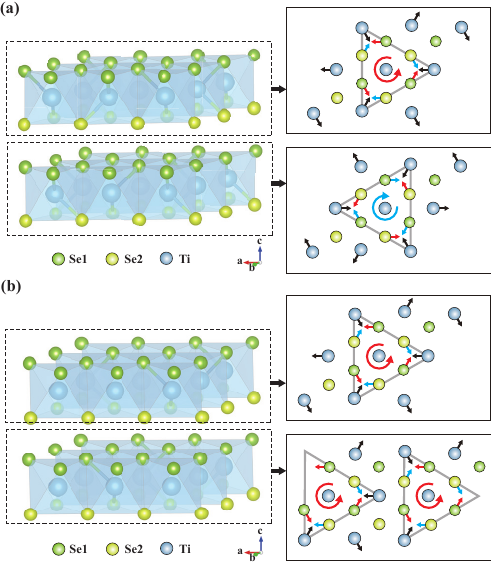}
  \caption{Atomic structure models for \textbf{(a)} $P\bar{3}c1$ structure and \textbf{(b)} similar model but for a newly proposed $C2$ structure. (Left panel)Three-dimensional view. (Right panel)Top view of atomic displacements (the red, blue, and black arrows) in a bilayer unit cell. Two dashed boxes in the left panel indicate the top and bottom layers. The helicity of Se1 (dark green circles) distortions is indicated by a circular arrow. The (anti)clockwise distortion is represented by blue(red) arrows. The Se-Ti-Se trimerizations are guided by the gray triangles (see details for atomic structures in FIG. S3(a) and (b) of Ref. \cite{SM2022}.).
  }
  \label{fig:3}
\end{figure}

\begin{figure}[tb!]
  \includegraphics{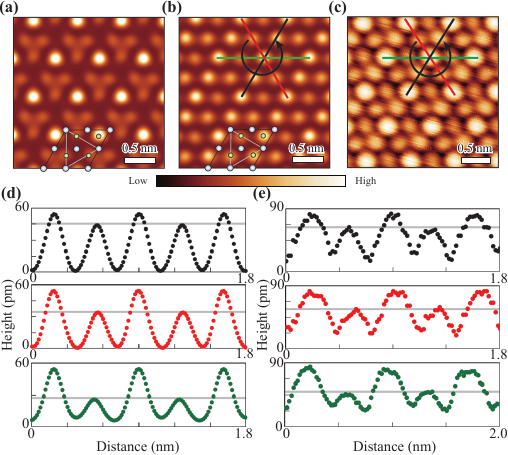}
  \caption{\textbf{(a),(b)} Simulated STM images for the $P\bar{3}c1$ and the $C2$ structure, respectively, which are proportional to the local density of states integrated between the Fermi energy and 30 meV in the empty states. The atomic positions in the topmost Se and Ti layers are indicated with the same format as FIG. \hyperref[fig:3]{3}. \textbf{(c)} The corresponding experimental STM image taken at a bias of $V_{s}$ = 100 meV. (see details in FIG. S1(c) and (d) of Ref. \cite{SM2022}.) \textbf{(d),(e)} The line profiles along three high symmetry directions indicated in (b) and (c), respectively. The simulation and the experiment are both for an anticlockwise domain.
}
  \label{fig:4}
\end{figure}

Since the chirality of the CDW was represented by the height differences of surface Se atoms, its microscopic origin has to be found in the CDW lattice distortion itself. In fact, we found only a weak reflection of the chirality in the scanning tunneling spectroscopy (STS) maps, supporting this initial thought (see STS data in FIG. S2 of Ref. \cite{SM2022}). The possible chiral atomic structures were then investigated by atomic relaxation within the DFT total energy calculations. The prevailing structure model of the $2\times2\times2$ CDW is the $P\bar{3}c1$ phase, which was introduced by the neutron diffraction measurements \cite{DiSalvo1976}. This is of course an achiral structure with a three-fold rotational symmetry. 
The CDW distortions of three Ti atoms are symmetric contraction towards the center but the Se atoms have helical distortions [Fig. 3(a)]. These helical distortions are compensated by the distortions in opposite directions in the neighboring layer of the out-of-plane of CDW unit cell to recover the rotational symmetry. That is, the bilayer is inversion symmetric. The STM simulation of this model shows no sign of chiral topography as shown in Fig. \hyperref[fig:3]{4(a)}, which is apparently not consistent with the STM observation. To resolve this discrepancy, we relaxed the atomic structure beyond the restriction of the three-fold rotational symmetry and allowed the translation within a bilayer. 
As a result, the $C2$ symmetry is found to be the energetically most favored structure as shown in FIG. \hyperref[fig:3]{3(b)} (see also FIG. S3 in Ref. \cite{SM2022}). We found that the energy gain of the $P\bar{3}c1$ ($C2$) phase from the pristine structure is 44 (52) meV per CDW unit cell. In fact, the possibility of this structure was hinted in the calculation of phonon instabilities of the original structure \cite{Subedi2022}. 
The CDW distortions within a single layer are very close in both models but the main difference is that the $C2$ structure has a translational shift of half atomic unit cell between the neighboring layer [FIG. \hyperref[fig:3]{3(b)}]. This interlayer translation breaks the inversion and glide mirror symmetry of the $P\bar{3}c1$ structure [FIG. \hyperref[fig:3]{3(a)}] and results in substantially larger out-of-plane distortions on the topmost layer. The resulting simulations shown in Fig. \hyperref[fig:4]{4(b)} and \hyperref[fig:4]{4(d)} reproduce the STM images excellently for the topmost layer chiral distortion. A marginal difference between the simulation and the experiment is the bias voltage; $V_{s}$ = 100 mV in the experiment but much closer to the Fermi level as 20 mV in the simulation. This can be caused by the different chemical potentials in the experiment due to impurities and unintentional dopants \cite{Spera2020} as well established in photoemission experiments \cite{Sugawara2016}.
Note further that the present model also resolves the issue in the electronic band structure of the $P\bar{3}c1$ model; the present model reproduces the direct band gap of the CDW phase as established well in photoemission experiments while the previous model cannot (see Fig. S3 in Ref. \cite{SM2022}). 

The present structure model successfully explains the chiral nature of the CDW from simply the structural degree of freedom within a layer and between the neighboring layer of a CDW unit cell. The present structure rules out the previous suggestion of an out-of-plane orbital ordering and the model of a weakly linked 1D chain in the out-of-plane direction \cite{Peng2022,VanWezelJasper2010}.
In addition, the previous STM study denied the chiral nature of the CDW mainly through the same chirality, if any, between the distortions of the neighboring layer \cite{Hildebrand2018}.
However, the present structure model does not require such a chiral difference between the layers.
Thus, those previous STM results can be consistently explained. On the other hand, the previous non-local measurements indicating the achiral nature of $\mathrm{TiSe_{2}}$ \cite{Lin2019,Chuang2020,Ueda2021,Zhang2022} can also be consistently incorporated by the presence of the two chiral domains and domain walls. In this respect, the global occurrence of an optically chiral phase induced by the cooling under chiral light can also straightforwardly be explained \cite{Xu2020}. The investigation of the microscopic mechanism of such an active interaction between the chiral domains and the light polarization would be an interesting topic for further studies.
The origin of the chiral CDW is simply structural where not only the in-plane but also the interlayer structural degree of freedom is important. Since the structural motive of the present chiral CDW distortion is widely shared in layered TMDC, the recent observations of chiral CDW in those materials can have similar origins.


\bibliography{1124_bib}


\end{document}


\title{Supplemental Material for \\``Microscopic Origin of Chiral Charge Density Wave in TiSe\texorpdfstring{$_{2}$}{2}''}

\author{Hyeonjung Kim}
\affiliation{Department of Physics, Pohang University of Science and Technology (POSTECH), Pohang 37673, Republic of Korea}
\affiliation{Center for Artificial Low Dimensional Electronic Systems, Institute for Basic Science (IBS), Pohang 37673, Republic of Korea}

\author{Kyung-Hwan Jin}
\affiliation{Center for Artificial Low Dimensional Electronic Systems, Institute for Basic Science (IBS), Pohang 37673, Republic of Korea}

\author{Han Woong Yeom}
\email{yeom@postech.ac.kr}
\affiliation{Department of Physics, Pohang University of Science and Technology (POSTECH), Pohang 37673, Republic of Korea}
\affiliation{Center for Artificial Low Dimensional Electronic Systems, Institute for Basic Science (IBS), Pohang 37673, Republic of Korea}

\maketitle

\begin{figure*}[t]
  \includegraphics{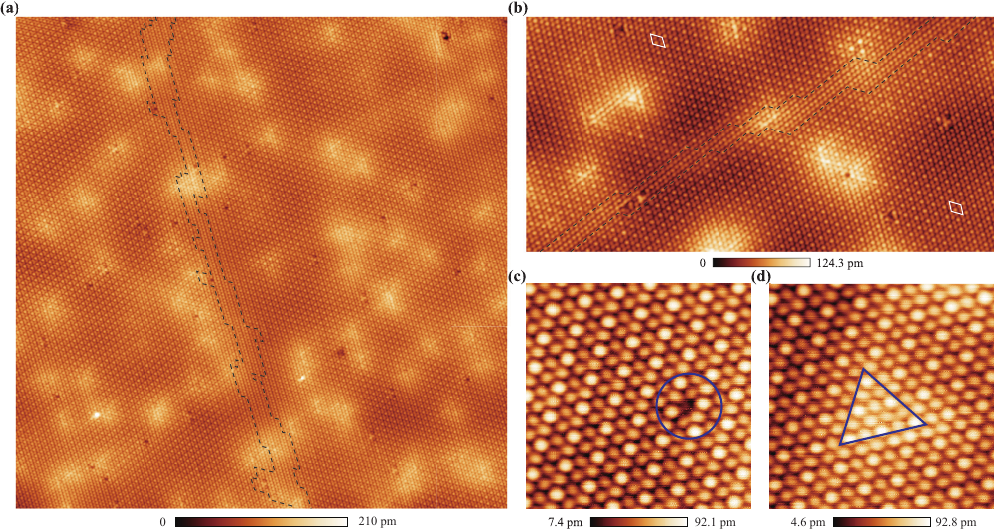}
  \caption{
  \textbf{(a)} The entire STM image (40 nm $\times$ 40 nm), where the the topographic STM image in FIG. 1(a) of the main text is extracted. \textbf{(b)} An STM image containing two chiral domains with a domain wall taken from a different area ($I_{t}$ = 100 pA, $V_{s}$ = 100 mV, and the size 30 nm $\times$ 30 nm). The white parallelograms represent the areas for the enlarged images in  the insets of FIG 2 of the main text are extracted (the upper right domain is a clockwise one). \textbf{(c),(d)} STM topographic images (5nm $\times$ 5nm) measured with the same tip condition as for the image in FIG. 4(c) at $I_{t}$ = 100 pA and $V_{s} = 100 mV$. The native defects correspond to a Se vacancy (c) and a Ti-intercalated site (d).\RaggedRight
  }
  \label{fig_s:1}
\end{figure*}

\begin{figure*}[t]
  \includegraphics{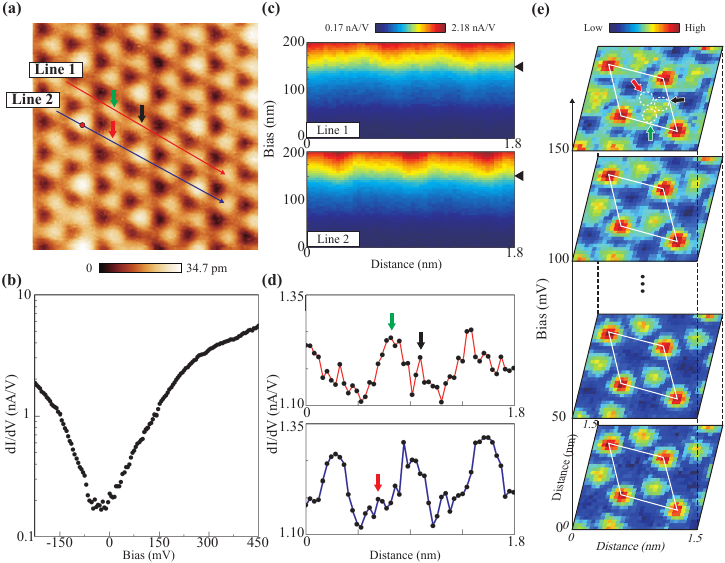}
  \caption{
   \textbf{(a)} STM topography (2.5 nm $\times$ 2.5 nm) measured with $V_{s}$ = 250 mV and $I_{t}$ = 100 pA. \textbf{(b)} Point STS (dI/dV) data at a red spot in (a). 
   \textbf{(c)} Empty-state line STS data along Line 1 and Line 2 in (a), which reveal the 2$\times$ period modulation in local density of states (LDOS) of the conduction band. 
   \textbf{(d)} The line profiles of the dI/dV intensity at 150 mV along Line 1 and Line 2 in (c) shows the 2$\times$ periodicity clearly. In a unit cell, the LDOS of three peaks (green, black, and red arrows measured on three different Se atoms within a unitcell as shown in (a)) have different intensities. 
   \textbf{(e)} The dI/dV maps at a few different biases in the empty state. The CDW unit cells are overlaid. The different LDOS at three different Se atoms are noticeable at the bias of 150 meV.
   For dI/dV measurements, the feedback loop was opened with the same start condition ($I_{t}$ = 100 pA and $V_{s}$ = 450 mV in (b) and (c); $I_{t}$ = 250 pA and $V_{s}$ = 400 mV in (e)) using a lock-in technique ((b) and (c) are recorded at frequency $f$ = 912 Hz and modulation $V_{m}$ = 15 mV; (e) is recorded at frequency $f$ = 1 kHz and modulation $V_{m}$ = 25 mV). \RaggedRight
  }
  \label{fig_s:2}
\end{figure*}

\begin{figure*}[t]
  \includegraphics{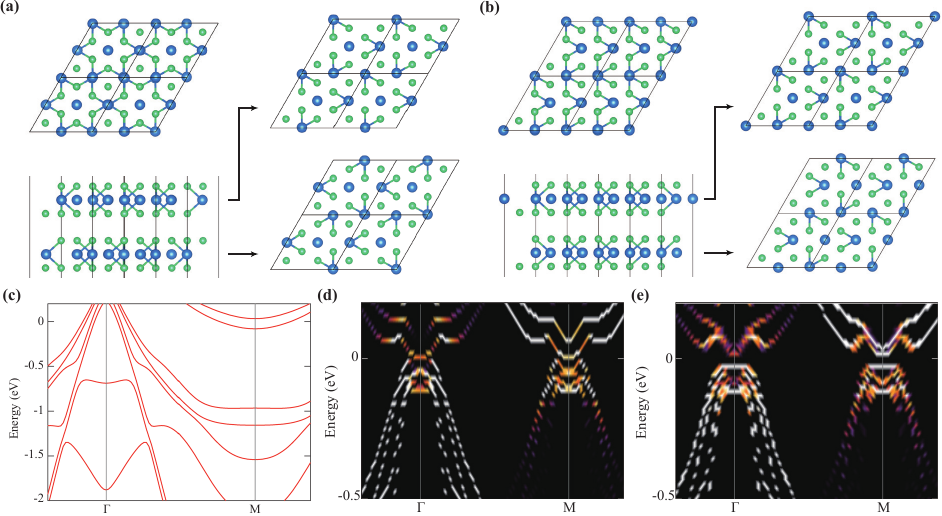}
   \caption{
  Atomic structures of a $1T\mathrm{-TiSe}_{2}$ bilayer in the CDW states with \textbf{(a)} $P\bar{3}c1$ and \textbf{(b)} $C2$ symmetry, respectively. The blue and green circles represent the Ti and Se atoms, respectively. In the top view of the bilayer with the $P\bar{3}c1$ structure, the trimerized Se-Ti-Se atoms are arranged into a a triangular shape.
  \textbf{(c)} Calculated band structure of the pristine $1T\mathrm{-TiSe}_{2}$ bilayer. The band structures of the CDW phases of the $1T\mathrm{-TiSe}_{2}$ bilayer with \textbf{(d)} $P\bar{3}c1$ and \textbf{(e)} $C2$ symmetry. Te band dispersions are unfoled into the Brillouin zone of the pristine unit cell to be compared with that of the pristine structure. While the $C2$ structure has a direct gap, the $P\bar{3}c1$ structure has an indirect gap at $M$ point.\RaggedRight
  }
  \label{fig_s:2}
\end{figure*}